\documentclass[12pt]{article}
\usepackage{amsfonts}
\usepackage{bm}
\begin{document}

\renewcommand{\theequation}{\thesection.\arabic{equation}}

\title{The Lorentz - invariant deformation of the Whitham
system for the non-linear Klein-Gordon equation.}

\author{A.Ya. Maltsev}

\date{
\centerline{L.D.Landau Institute for Theoretical Physics,}
\centerline{119334 ul. Kosygina 2, Moscow, maltsev@itp.ac.ru}}

\maketitle

\begin{abstract}
We consider the deformation of the Whitham system for the non-linear 
Klein-Gordon equation having the Lorentz-invariant form. Using the
Lagrangian formalism of the initial system we obtain the first
non-trivial correction to the Whitham system in the Lorentz-invariant
approach. 
\end{abstract}

\section{Introduction.}

We will consider the deformation of the Whitham system for the
non-linear Klein-Gordon equation

\begin{equation}
\label{nKG}
\varphi_{tt} \,\, - \,\, \varphi_{xx} \,\, + \,\,
V^{\prime}(\varphi) \,\,\, = \,\,\, 0
\end{equation}

 The Whitham method is connected with the slow modulations of
the periodic or quasiperiodic $m$-phase solutions of the nonlinear
systems

\begin{equation}
\label{insyst}
F^{i}(\bm{\varphi}, \bm{\varphi}_{t}, \bm{\varphi}_{x}, \dots ) 
\,\, = \,\, 0
\,\,\,\,\,\,\,\, , \,\,\,\,\, i = 1, \dots, n \,\,\, , \,\,\,
\bm{\varphi} = (\varphi^{1}, \dots, \varphi^{n})
\end{equation}
which are represented usually in the form

\begin{equation}
\label{phasesol}
\varphi^{i} (x,t) \,\, = \,\, \Phi^{i} \left( {\bf k}({\bf U})\, x
\, + \, \bm{\omega}({\bf U})\, t \, + \, \bm{\theta}_{0}, \,
{\bf U} \right)
\end{equation}

 In these notations the functions ${\bf k}({\bf U})$ and
$\bm{\omega}({\bf U})$ play the role of the "wave numbers"
and "frequencies" of $m$-phase solutions and $\bm{\theta}_{0}$
are the initial phase shifts.

 The functions $\Phi^{i}(\bm{\theta})$ satisfy the system

\begin{equation}
\label{phasesyst}
F^{i} \left( {\bf \Phi}, \omega^{\alpha} {\bf \Phi}_{\alpha},
k^{\beta} {\bf \Phi}_{\beta}, \dots \right) \,\,\, \equiv 
\,\,\, 0
\,\,\,\,\,\,\,\, , \,\,\,\,\, i = 1, \dots, n
\end{equation}
and we choose (in a smooth way) some function 
${\bf \Phi}(\bm{\theta}, {\bf U})$ for every ${\bf U}$ as having
"zero initial phase shifts". The full set of $m$-phase solutions
of (\ref{insyst}) can then be represented in the form 
(\ref{phasesol}).

 For the $m$-phase solutions of (\ref{insyst}) we have then
${\bf k}({\bf U}) = (k^{1}({\bf U}), \dots, k^{m}({\bf U}))$,
$\bm{\omega}({\bf U}) = (\omega^{1}({\bf U}), \dots, 
\omega^{m}({\bf U}))$, 
$\bm{\theta}_{0} = (\theta^{1}, \dots, \theta^{m})$,
where ${\bf U}= (U^{1}, \dots, U^{N})$ are the parameters
of the solution. We require also that all the functions
$\Phi^{i}(\bm{\theta}, {\bf U})$ are $2\pi$-periodic with
respect to every $\theta^{\alpha}$, $\alpha = 1, \dots, m$.

 For one-phase solutions of (\ref{nKG}) we have obviously

\begin{equation}
\label{phsystnKG}
(\omega^{2} - k^{2}) \Phi_{\theta\theta} \,\, + \,\,
V^{\prime}\left( \Phi \right) \,\,\, = \,\,\, 0
\end{equation}
where $\Phi(\theta, \omega^{2} - k^{2})$ is $2\pi$-periodic
function of $\theta$.

 In Whitham approach (see \cite{whith1,whith2,whith3,luke,AblBenny,
Abl1,Abl2,Hayes,gurpit1,gurpit2,dm,theorsol,dn1,LaxLev,Nov,AvNov,
gurpit3,AvKrichNov,krichev1,Haberman,dn2,dn3}) the parameters 
${\bf U}$ become the slow functions of $x$ and $t$:
${\bf U} = {\bf U}(X,T)$, where $X = \epsilon x$, $T = \epsilon t$
($\epsilon \rightarrow 0$).

 More precisely (see \cite{luke}), we try to find the asymptotic 
solutions

\begin{equation}
\label{whithsol}
\varphi^{i}(\bm{\theta}, X, T) \,\,\, = \,\,\, \sum_{k\geq0} 
\Psi^{i}_{(k)} \left( {{\bf S}(X,T) \over \epsilon} + 
\bm{\theta}_{0}, \, X, \, T \right) \,\, \epsilon^{k}
\end{equation}
(where all $\bm{\Psi}_{(k)}$ are $2\pi$-periodic in $\bm{\theta}$)
which satisfy the system (\ref{insyst}), i.e.

$$F^{i} \left( \bm{\varphi}, \epsilon \bm{\varphi}_{T}, 
\epsilon \bm{\varphi}_{X}, \dots \right) \,\,\, = \,\,\, 0 
\,\,\,\,\,\,\,\, , \,\,\,\,\, i = 1, \dots, n $$

 The function ${\bf S}(X,T) = (S^{1}(X,T), \dots, S^{m}(X,T))$
is the "modulated phase" of the solution (\ref{whithsol}) and the
function $\bm{\Psi}_{(0)}(\bm{\theta}, X, T)$ belongs at every $X$
and $T$ to the family of $m$-phase solutions of (\ref{insyst}).
We have then

\begin{equation}
\label{psi0}
\bm{\Psi}_{(0)} (\bm{\theta},X,T) \,\,\, = \,\,\,
\bm{\Phi} \left( \bm{\theta} + \bm{\theta}_{0}(X,T), {\bf U}(X,T)
\right)
\end{equation}
and

$$S^{\alpha}_{T}(X,T) \, = \, \omega^{\alpha}({\bf U}) \,\,\, ,
\,\,\,\,\, S^{\alpha}_{X}(X,T) \, = \, k^{\alpha}({\bf U}) $$
as follows from the substitution of (\ref{whithsol}) in the system
(\ref{insyst}).

 The functions $\bm{\Psi}_{(k)} (\bm{\theta},X,T)$ are defined
from the linear systems

\begin{equation}
\label{ksyst}
{\hat L}^{i}_{j[{\bf U}, \bm{\theta}_{0}]}(X,T) \,\,
\Psi_{(k)}^{j} (\bm{\theta},X,T) \,\,\, = \,\,\,
f_{(k)}^{i} (\bm{\theta},X,T)
\end{equation}  
where ${\hat L}^{i}_{j[{\bf U}, \bm{\theta}_{0}]}(X,T)$
is a linear operator given by the linearizing of the system
(\ref{phasesyst}) on the solution (\ref{psi0}). The resolvability
conditions of the systems (\ref{ksyst}) can be written as the
orthogonality conditions of the functions
${\bf f}_{(k)} (\bm{\theta},X,T)$ to all the 
"left eigen vectors" (the eigen vectors of adjoint operator)
$\bm{\kappa}^{(q)}_{[{\bf U}(X,T)]}
(\bm{\theta} + \bm{\theta}_{0}(X,T))$ of the operator
${\hat L}^{i}_{j[{\bf U}, \bm{\theta}_{0}]}(X,T)$ corresponding to
zero eigen-values. The resolvability conditions of (\ref{ksyst})
for $k = 1$ together with
 
$$k^{\alpha}_{T} = \omega^{\alpha}_{X} $$
give the Whitham system for $m$-phase solutions of (\ref{insyst})
playing the central role in the slow modulations approach.

 We have for the equation (\ref{nKG})

$${\hat L}_{[k,\omega,\theta_{0}]} \,\,\, = \,\,\,
(\omega^{2} - k^{2}) {\partial^{2} \over \partial \theta^{2}} 
\, + \,  V^{\prime\prime} \left(\Phi(\theta + \theta_{0}, k, \omega) 
\right)$$
and the function $\Phi_{\theta}(\theta + \theta_{0}, k, \omega)$
plays the role of both "left" and "right" eigen-vector corresponding
to zero eigen-value.

 The function $f_{(1)} (\theta, X, T)$ is defined as

\begin{equation}
\label{newf1}
f_{(1)} \,\, = \,\, - S_{TT} \Phi_{\theta} \, - \,
2 S_{T} \Phi_{\theta T} \, + \, S_{XX} \Phi_{\theta} \, + \,
2 S_{X} \Phi_{\theta X}
\end{equation}
and the Whitham system can be written as

\begin{equation}
\label{whithnKG}
\left( \omega \int_{0}^{2\pi} \Phi_{\theta}^{2} \,
{d \theta \over 2\pi} \right)_{T} \,\,\, = \,\,\,
\left( k \int_{0}^{2\pi} \Phi_{\theta}^{2} \,
{d \theta \over 2\pi} \right)_{X}
\end{equation}
$$k_{T} \,\,\, = \,\,\, \omega_{X} $$
(\cite{whith1,whith2,whith3,luke}).

 The Whitham system is a so-called system of Hydrodynamic Type,
which can be written in the form

\begin{equation}
\label{ABsyst}
A^{\nu}_{\mu} ({\bf U}) \, U^{\mu}_{T} \,\,\, = \,\,\,
B^{\nu}_{\mu} ({\bf U}) \, U^{\mu}_{X}
\end{equation}
with some matrices $A({\bf U})$ and $B({\bf U})$. In generic case
the system (\ref{ABsyst}) can be resolved w.r.t. the time 
derivatives of ${\bf U}$ and written in the evolution form

\begin{equation}
\label{HTsyst}
U^{\nu}_{T} \,\,\, = \,\,\, V^{\nu}_{\mu} ({\bf U}) \, U^{\mu}_{X}
\,\,\,\,\,\,\,\, , \,\,\,\,\, \nu = 1, \dots, N
\end{equation}
(where $V = A^{-1} B$).
 
 The Lagrangian properties of the Whitham system were investigated
by Whitham (\cite{whith3}) who suggested also the method of 
"averaging" of the Lagrangian function to get the  Lagrangian 
function for the Whitham system.

 Another important procedure is the procedure of "averaging"
of local Hamiltonian structures suggested by B.A. Dubrovin and
S.P. Novikov (\cite{dn1,dn2,dn3}). The Dubrovin - Novikov
procedure gives a field-theoretical Hamiltonian structure
of Hydrodynamic Type for the system (\ref{HTsyst}) with a
Hamiltonian function having the hydrodynamic form
$H = \int h({\bf U}) dX$. Dubrovin - Novikov bracket for the system 
(\ref{HTsyst}) has the form 

\begin{equation}
\label{DNbr}
\{U^{\nu}(X), U^{\mu}(Y)\} \,\, = \,\, g^{\nu\mu}({\bf U}) \,
\delta^{\prime}(X-Y) \, + \, b ^{\nu\mu}_{\lambda}({\bf U}) \,
U^{\lambda}_{X} \, \delta (X-Y)
\end{equation}
which is called also the local Poisson bracket of Hydrodynamic Type.

 The Hamiltonian properties of the systems (\ref{HTsyst}) are 
strongly correlated with their integrability properties. Thus it
was proved by S.P. Tsarev (\cite{Tsarev}) that all the
diagonalizable systems (\ref{HTsyst}) having Dubrovin - Novikov
Hamiltonian structure can in fact be integrated (S.P. Novikov
conjecture). Actually the same is true also for the
diagonalizable systems (\ref{HTsyst}) having more general
weakly-nonlocal Mokhov-Ferapontov or Ferapontov Hamiltonian
structures (\cite{mohfer1,fer1,fer2,fer3,fer4,pavlov3,PhysD}).
Let us mention here that the generalization of Dubrovin - Novikov
procedure for the weakly-nonlocal case was suggested in
\cite{IJMMS}.

 The theory of the systems and hierarchies (\ref{HTsyst}) having
the bi-Hamiltonian formalism plays the basic role in the theory
of Frobenius manifolds 
(\cite{Dubrov1,Dubrov2,Dubrov3,Dubrov4,Dubrov5})
giving the classification of of Topological Quantum Fields
Theories based on WDVV-equation. Let us say here that 
Dubrovin - Novikov procedure gives also the possibility of
construction of bi-Hamiltonian formalism of the Whitham system
(\ref{HTsyst}) if the initial system (\ref{insyst}) has a
local field-theoretical bi-Hamiltonian formalism. As a corollary
of this fact, the Whitham hierarchies corresponding to the most
of the integrable systems (like KdV or NLS) give in fact the
important examples of Frobenius manifolds.

 The higher corrections to Topological Quantum Field theories
require the deformations (\cite{DubrZhang1,DubrZhang2,DubrZhang3})
of the Hydrodynamic Type hierarchies (\ref{HTsyst}) having the
form

\begin{equation}
\label{defsyst}
U^{\nu}_{T} \,\,\, = \,\,\, V^{\nu}_{\mu} ({\bf U}) \, U^{\mu}_{X}
\,\, + \,\, \sum_{k\geq2} v_{(k)}^{\nu} ({\bf U}, {\bf U}_{X},
\dots, {\bf U}_{kX}) \,\, \epsilon^{k-1}
\end{equation}
where all $v_{(k)}^{\nu}$ are smooth functions polynomial in
the derivatives ${\bf U}_{X}$, $\dots$, ${\bf U}_{kX}$ and
having degree $k$ according to the following gradation rule:

1) All the functions $f({\bf U})$ have degree $0$;

2) The derivatives $U^{\nu}_{kX}$ have degree $k$;

3) The degree of the product of two functions having certain 
degrees is equal to the sum of their degrees.

 The deformation (\ref{defsyst}) of the system (\ref{HTsyst})
implies also the deformation of the corresponding 
(bi-)Hamiltonian structures (\ref{DNbr})

$$\{U^{\nu}(X), U^{\mu}(Y)\} \,\, = \,\, 
\{U^{\nu}(X), U^{\mu}(Y)\}_{0} \,\, + $$
\begin{equation}
\label{defbr}
+ \,\, \sum_{k\geq2}
\epsilon^{k-1} \, \sum_{s=0}^{k} 
B^{\nu\mu}_{(k)s}({\bf U}, {\bf U}_{X}, \dots, {\bf U}_{(k-s)X})
\,\, \delta^{(s)} (X - Y)
\end{equation}
where all $B^{\nu\mu}_{(k)s}$ are polynomial w.r.t.
derivatives ${\bf U}_{X}$, $\dots$, ${\bf U}_{(k-s)X}$ and
have degree $(k-s)$.
 
 The deformation (\ref{defsyst}) of the hierarchy of Hydrodynamic
Type with the deformation (\ref{defbr}) of the corresponding
bi-Hamiltonian structure plays the basic role in the procedure
of deformation of Frobenius manifold 
(\cite{DubrZhang1,DubrZhang2,DubrZhang3}). Let us say here that the
general theory of deformations of integrable hierarchies 
of Hydrodynamic Type as well as the bi-Hamiltonian structures
of Dubrovin - Novikov type is being actively studied by now.
Let us also point out here the recent papers where the important 
results in this area were obtained 
\cite{Lorenzoni,LiuZhang1,LiuZhang2,DubrLiuZhang,DubrZhangZuo}.
Let us call the deformation (\ref{defsyst}) of any system
of Hydrodynamic Type the Dubrovin - Zhang deformation.

 As was first pointed out in \cite{AblBenny} the higher 
corrections in Whitham method satisfy the more complicated equations
including "dispersive terms" and the Whitham system (\ref{ABsyst})
should in fact contain also the higher derivatives ("dispersion")
being considered in the next orders of $\epsilon$.\footnote{Also
the multi-phase Whitham method was first considered in 
\cite{AblBenny,Abl1,Abl2}.} 

 In \cite{deform} the general procedure of deformation of
Hyperbolic Whitham systems based on the "renormalization" of
parameters was suggested. The procedure suggested in \cite{deform}
gives the deformation of the Whitham system (\ref{HTsyst}) having
the Dubrovin - Zhang form (\ref{defsyst}). This method can be
considered from our point of view as the solution of the first
part of the problem set by B.A. Dubrovin and connected with the
deformations of Frobenius manifolds. Namely, B.A. Dubrovin problem
requires the construction of the deformation of the Whitham system
(\ref{HTsyst}) in Dubrovin - Zhang form and the corresponding
bracket (\ref{DNbr}) also having Dubrovin - Zhang form (\ref{defbr}).
The solution of Dubrovin problem for the case of bi-Hamiltonian
integrable systems gives in fact the deformation of Frobenius
manifolds defined by the corresponding Whitham hierarchies which
are bi-Hamiltonian according to Dubrovin - Novikov procedure.
Let us say here also that in \cite{deflag} the Lagrangian properties
of the deformations of Whitham systems suggested in \cite{deform}
were investigated. However, the generalization of Dubrovin - Novikov 
procedure giving the deformations of the "averaged" brackets
(\ref{DNbr}) was not considered yet.

\section{Lorentz-invariant deformation of the Whitham 
\newline system.}
\setcounter{equation}{0}

 In this paper we will consider the deformation of the Whitham 
system for the equation (\ref{nKG}) having Lorentz-invariant
form. Like in \cite{deform,deflag} we will use here the 
"renormalization" of parameters of one-phase solutions of
(\ref{nKG}) after the construction of the solution 
(\ref{whithsol}). Namely, we introduce the "renormalized"
modulated phase

$$S(X, T, \epsilon) \,\,\, = \,\,\, \sum_{k\geq0}
S_{(k)} (X, T) \,\, \epsilon^{k} $$
which is the "physically observable" quantity and make then the
"re-expansion" of the series (\ref{whithsol}) using the higher
derivatives of $S(X, T, \epsilon)$ instead of the parameter 
$\epsilon$. The function

\begin{equation}
\label{mainappr}
\Phi \left( {S(X,T,\epsilon) \over \epsilon} \, + \, \theta,
\, S_{X}(X,T,\epsilon), \, S_{T}(X,T,\epsilon) \right)
\end{equation}
will play now the role of the main approximation in the
"renormalized" expansion while all the higher terms will contain
the higher derivatives of the function $S(X, T, \epsilon)$.
The terms of the new expansion will now be $\epsilon$-dependent
and constructed according to the new "gradation rule" put on the
derivatives of $S(X, T, \epsilon)$.

 In this situation we can omit in fact the "unobservable"
parameter $\epsilon$ and use just the functionals of the
"renormalized" modulated phase $S(X, T) = S(X, T, \epsilon)$.
Following \cite{deform,deflag} let us then omit the parameter
$\epsilon$ in all calculations (or put formally $\epsilon = 1$),
however, we will keep the notations $X$ and $T$ for the spatial
and time variables just to emphasize that the parameters of
one-phase solution are the slow functions of $x$ and $t$. Like
in \cite{deform,deflag} we define now the "renormalization rule"
which determines the renormalized phase $S(X, T, \epsilon)$.
Namely, we look now for the solution of (\ref{nKG}) having
the form

\begin{equation}
\label{newexp}
\varphi(\theta, X, T) \,\,\, = \,\,\,
\Phi \left( S(X,T) + \theta, S_{X}, S_{T} \right) \, + \,
\sum_{k\geq1} {\tilde \Psi}_{(k)} 
\left( S(X,T) + \theta, X, T \right)
\end{equation}
where the functions ${\tilde \Psi}_{(k)}$ are the local 
functionals of $S(X,T)$ and it's derivatives having gradation 
degree $k$ (defined below). All the functions 
${\tilde \Psi}_{(k)}(\theta,X,T)$ are defined from the linear
systems

\begin{equation}
\label{newksyst}
{\hat L}_{[S_{X},S_{T}]} {\tilde \Psi}_{(k)}(\theta,X,T)
\,\,\, = \,\,\, {\tilde f}_{(k)} (\theta,X,T)
\end{equation}
where

$${\hat L}_{[S_{X},S_{T}]} \,\,\, = \,\,\, \left(
S_{T}^{2} \, - \, S_{X}^{2} \right) \, 
{\partial^{2} \over \partial \theta^{2}} \,\, + \,\,
V^{\prime\prime} \left( \Phi(\theta, S_{X}, S_{T}) \right) $$
is the linearization of the equation (\ref{phsystnKG}) on the 
function $\Phi(\theta, S_{X}, S_{T})$
and ${\tilde f}_{(k)} (\theta,X,T)$ is the discrepancy having
gradation degree $k$. The system (\ref{newksyst}) defines the
function ${\tilde \Psi}_{(k)}(\theta,X,T)$ modulo the function

$$c(X,T) \, \Phi_{\theta}(\theta, S_{X}, S_{T}) $$
which belongs to the kernel of the operator 
${\hat L}_{[S_{X},S_{T}]}$ at every $X$ and $T$. We require
now that the function $S(X,T)$ is normalized in the "optimal"
way, such that

\begin{equation}
\label{normcond}
\int_{0}^{2\pi} \Phi_{\theta} (\theta, S_{X}, S_{T}) \,\,
{\tilde \Psi}_{(k)}(\theta,X,T) \,\, {d \theta \over 2\pi} 
\,\,\, = \,\,\, 0
\end{equation}
for all $X$ and $T$.

 The condition (\ref{normcond}) defines now uniquely the
function ${\tilde \Psi}_{(k)}(\theta,X,T)$ satisfying the
system (\ref{newksyst}).

 Let us speak now about the "gradation rule" we are going to 
use here.

 Let us say first that the gradation used in \cite{deform,deflag}
and giving the deformation of the Whitham system in
Dubrovin - Zhang form was defined by the $X$-derivatives of
the parameters ${\bf U}$ of $m$-phase solutions (\ref{phasesol}).
Namely, in \cite{deform,deflag} the following gradation rule
was used:

1) The functions $f(\omega, k) = f(S_{T}, S_{X})$ have degree
$0$;

2) The derivatives $\omega_{kX}$, $k_{kX}$ have degree $k$;

3) The degree of the product of two functions having certain 
degrees is equal to the sum of their degrees.

 According to this gradation rule the higher $T$-derivatives
of the parameters $(\omega, k)$ do not have certain degrees
and it is required just that the functions $\omega_{kT}$, 
$k_{kT}$ can be represented as the (infinite) series of terms
having degree $\geq k$. The expression of $(k_{T}, \omega_{T})$
in terms of $X$ derivatives of the functions $(k, \omega)$

$$k_{T} \,\,\, = \,\,\, \omega_{X}$$
\begin{equation}
\label{defDZ}
\omega_{T} \,\,\, = \,\,\, \sum_{k\geq1} \sigma_{(k)}
(k, \omega, k_{X}, \omega_{X}, \dots )
\end{equation}
plays then the role of the deformation of the Whitham system 
having Dubrovin - Zhang form. All the functions
$\sigma_{(k)}(k, \omega, k_{X}, \omega_{X}, \dots )$ have 
degree $k$ and are defined from the compatibility conditions

\begin{equation}
\label{ortcond}
\int_{0}^{2\pi} \Phi_{\theta} (\theta, S_{X}, S_{T}) \,\,
{\tilde f}_{(k)}(\theta,X,T) \,\, {d \theta \over 2\pi} 
\,\,\, = \,\,\, 0
\end{equation}
for the system (\ref{newksyst}) in the $k$-th order.
The system (\ref{defDZ}) plays then the role of the 
"full Whitham system" including all the orders of asymptotic 
expansion.

 In this paper, however, we have to use some Lorentz-invariant
gradation rule to provide the Lorentz-invariant deformation
of the Whitham system (\ref{whithnKG}).

 Let us introduce now two differential operators

$${\hat \xi}_{1} \,\, = \,\, S_{T} {\partial \over \partial T}
\, - \, S_{X} {\partial \over \partial X}  \,\,\,\,\,\,\,\, ,
\,\,\,\,\,\,\,\,
{\hat \xi}_{2} \,\, = \,\, S_{X} {\partial \over \partial T}
\, - \, S_{T} {\partial \over \partial X} $$

 Easy to see that both ${\hat \xi}_{1}$ and ${\hat \xi}_{2}$
are Lorentz-invariant. Let us use notation 
$\mu = S_{T}^{2} - S_{X}^{2}$. We have

$${\hat \xi}_{1} S \,\, = \,\, S_{T}^{2} - S_{X}^{2} 
\,\, = \,\, \mu ,\,\,\,\,\,\,\, ,   
\,\,\,\,\,\,\,\, {\hat \xi}_{2} S \,\, = \,\, 0 $$
 
 We put by definition $deg \, f(\mu) = 0$ for any smooth function
$f(\mu)$. Easy to see then that the expressions

$${\hat \xi}_{1} \mu \,\, = \,\, 2 S_{T}^{2} S_{TT} \, + \,
2 S_{X}^{2} S_{XX} \, - \, 4 S_{T} S_{X} S_{XT} $$
and

$${\hat \xi}_{2} \mu \,\, = \,\, 2 S_{T} S_{X} (S_{TT} + S_{XX})
\, - \, 2 (S_{T}^{2} + S_{X}^{2}) S_{XT} $$
are both Lorentz-invariant functions. We then define the general
function having degree $1$ in the form

$$f \left[ S \right] \,\,\, = \,\,\, 
a (\mu) \, {\hat \xi}_{1} \mu \, + \, 
b (\mu) \, {\hat \xi}_{2} \mu $$

 Let us note that the function 
$\nu \, = \, S_{TT} \, - \, S_{XX}$ does not have certain degree
in this approach and will be represented as the infinite sum of
the terms having degree $\geq 1$. 

 Let us define now the functions
with higher degrees using the operator ${\hat \xi}_{1}$. Namely,
we introduce the family ${\cal M}$ of functions with certain 
gradation degrees using the following rule:

\vspace{0.5cm}

1) All the smooth functions $f(\mu)$ belong to ${\cal M}$
and have degree $0$;

2) The functions $Q = {\hat \xi}_{1} \mu$, 
$P = {\hat \xi}_{2} \mu$ belong to ${\cal M}$ and
$deg \, Q \, = \, deg \, P \, = \, 1$.

3) If the functions $F$ and $G$ belong to ${\cal M}$ then
$FG$ belongs to ${\cal M}$ and

$$deg \, (FG) \,\,\, = \,\,\, deg \, F \, + \, deg \, G $$

4) If the function $F$ belongs to ${\cal M}$ then the function
${\hat \xi}_{1} F$ also belongs to ${\cal M}$ and

$$deg \, ({\hat \xi}_{1} F) \,\,\, = \,\,\, deg \, F \, + \, 1$$

\vspace{0.5cm}

 Easy to see that all the functions from the family ${\cal M}$
are Lorentz-invariant. 

\vspace{0.5cm}

 The function $\nu \, = \, S_{TT} \, - \, S_{XX}$ does not
belong to the family ${\cal M}$ and we do not prescribe any certain
degree to this function. Instead, we put now the conditions

\begin{equation}
\label{defLor}
S_{TT} \, - \, S_{XX} \,\,\, = \,\,\, \sum_{k\geq1} \nu_{(k)}
\end{equation}
where every $\nu_{(k)}$ is represented as a sum functions belonging
to ${\cal M}$ and having degree $k$.

 The system (\ref{defLor}) will now play the role of the deformation
of the Whitham system in the Lorentz-invariant representation. Every
function $\nu_{(k)}$, $k\geq1$ is defined now from the compatibility
conditions (\ref{ortcond}) of the system (\ref{newksyst}) in the
$k$-th order.

 The equation

$$S_{TT} \, - \, S_{XX} \,\,\, = \,\,\, \nu_{(1)} $$
coincides with the Whitham system (\ref{whithnKG}) and we have

\begin{equation}
\label{nu1}
\nu_{(1)} \,\,\, = \,\,\, - \left. \left[ \left( 
\omega {\partial \over \partial T} - 
k {\partial \over \partial X} \right) W(\mu) \right] \right/
W(\mu) \,\,\, = \,\,\, 
- \left({\hat \xi}_{1} W(\mu) \right) \left/ W(\mu) \right.
\end{equation}
where
$$W(\mu) \,\,\, = \,\,\, \int_{0}^{2\pi} 
\Phi_{\theta}^{2}(\theta, \mu) \,\, {d \theta \over 2\pi} $$

\vspace{0.5cm}

 Let us come back to the expansion (\ref{newexp}). We will
assume now that every function 
${\tilde \Psi}_{(k)}(\theta, X, T)$ is represented by a sum of 
functions which belong to the family ${\cal M}$ (at every $\theta$)
and have degree $k$. For the function

\begin{equation}
\label{phifunc}
\phi (\theta, X, T) \,\,\, = \,\,\, 
\varphi (\theta - S(X,T), X, T) \,\,\, = \,\,\,
\Phi (\theta, X, T) \, + \, \sum_{k\geq1} 
{\tilde \Psi}_{(k)}(\theta, X, T)
\end{equation}
we have the equation

$$\left(S_{T}^{2} - S_{X}^{2} \right) \phi_{\theta\theta}
\, + \, V^{\prime} (\phi) \, + \, 2 \left(
S_{T} {\partial \over \partial T} - 
S_{X} {\partial \over \partial X} \right) \phi_{\theta} \, + \,
\left( S_{TT} - S_{XX} \right) \phi_{\theta} \, + \,
\phi_{TT} \, - \, \phi_{XX} \,\,\, = \,\,\, 0 $$

 It's not difficult to check the relation

$${\partial^{2} \over \partial T^{2}} \, - \,
{\partial^{2} \over \partial X^{2}} \,\,\, = \,\,\, - \,
{\hat \xi}_{1}^{+} \, {1 \over \mu} \, {\hat \xi}_{1} \, + \,
{\hat \xi}_{2}^{+} \, {1 \over \mu} \, {\hat \xi}_{2} \,\,\, = $$
$$= \,\,\, {1 \over \mu} \left( {\hat \xi}_{1}^{2} -
{\hat \xi}_{2}^{2} \right) \, - \, {1 \over \mu^{2}}
\left( {\hat \xi}_{1} \mu \right) {\hat \xi}_{1} \, + \,
{1 \over \mu^{2}} \left( {\hat \xi}_{2} \mu \right) {\hat \xi}_{2}
\, + \, {1 \over \mu} \left( S_{TT} - S_{XX} \right) {\hat \xi}_{1}$$
where

$${\hat \xi}_{1}^{+} \,\, = \,\, - \, 
{\partial \over \partial T} S_{T} \, + \,
{\partial \over \partial X} S_{X} \,\,\,\,\,\,\,\, ,
\,\,\,\,\,\,\,\, {\hat \xi}_{2}^{+} \,\, = \,\, - \, {\hat \xi}_{2}$$

 We have then

$$\mu \, \phi_{\theta\theta} \, + \, V^{\prime} (\phi) \, + \,
2 \, {\hat \xi}_{1} \, \phi_{\theta} \, + \,
\left( S_{TT} - S_{XX} \right) \phi_{\theta} \, + $$
$$+ \, {1 \over \mu} 
\left( {\hat \xi}_{1}^{2} - {\hat \xi}_{2}^{2} \right)
\phi \, - \, {1 \over \mu^{2}} \left( {\hat \xi}_{1} \mu \right)
\left( {\hat \xi}_{1} \phi \right) \, + \,
{1 \over \mu^{2}} \left( {\hat \xi}_{2} \mu \right)
\left( {\hat \xi}_{2} \phi \right) \, + \, {1 \over \mu} \,
\left( S_{TT} - S_{XX} \right) \left( {\hat \xi}_{1} \phi \right)
\,\, = \,\, 0 $$

 Using the relation (\ref{defLor}) we can write then in the
$k$-th order:

$$\mu \, {\tilde \Psi}_{(k)\theta\theta} \, + \,
V^{\prime\prime} \left( \Phi \right) {\tilde \Psi}_{(k)} \,\, = \,\,
- \sum_{l=1}^{k} \nu_{(l)} \left( {\tilde \Psi}_{(k-l)\theta}
\, + \, {1 \over \mu} \, {\hat \xi}_{1} \, 
{\tilde \Psi}_{(k-l-1)} \right) \, - \, {1 \over \mu} \left(
{\hat \xi}_{1}^{2} {\tilde \Psi}_{(k-2)} \right) \, + $$
\begin{equation}
\label{xiexpr}
+ \, {1 \over \mu^{2}} \left( {\hat \xi}_{1} \mu \right) \left(
{\hat \xi}_{1} {\tilde \Psi}_{(k-2)} \right) \, - \, 
2 \, {\hat \xi}_{1} \, {\tilde \Psi}_{(k-1)\theta} \, + \,
{1 \over \mu} \left({\hat \xi}_{2}^{2} \phi \right)^{[k]} \, - \,
{1 \over \mu^{2}} \left( {\hat \xi}_{2} \mu \right) 
\left( {\hat \xi}_{2} \phi \right)^{[k-1]} \, - \,
V^{\prime}_{(k)}
\end{equation}
(where ${\tilde \Psi}_{(q)} \equiv 0$ for $q < 0$,
and we put also 
${\tilde \Psi}_{(0)} \equiv \Phi(\theta,S_{X},S_{T})$).

 The expressions $({\hat \xi}_{2}^{2} \phi )^{[k]}$ and
$({\hat \xi}_{2} \phi )^{[k-1]}$ represent here the terms
of order $k$ and $k-1$
of the gradated expansions of ${\hat \xi}_{2}^{2} \phi$ and
${\hat \xi}_{2} \phi$ respectively and $V^{\prime}_{(k)}$ is a sum
of functions having degree $k$ given by the expansion of 
$V^{\prime}$ (except $V^{\prime\prime}(\Phi) {\tilde \Psi}_{(k)}$).

 To use the equation (\ref{xiexpr}) we have to define now the
action of the operator ${\hat \xi}_{2}$ on the family ${\cal M}$.
In general, the expressions ${\hat \xi}_{2} F$, $F \in {\cal M}$
do not have certain degrees according to our gradation rule, so we
have to represent these expressions as the infinite series of terms
having certain degrees. We will assume naturally that the expansion
of ${\hat \xi}_{2} F$ will always start with a term of degree
$k+1$ if $deg \, F = k$. Let us prove now the following Lemma:

\vspace{0.5cm}

{\bf Lemma 2.1.}

{\it The relation (\ref{defLor}) defines uniquely the gradated
expansion of any expression $({\hat \xi}_{2} F)$ where $F$ is a
function having certain degree.}

\vspace{0.5cm}

Proof.

We have by definition that the functions $({\hat \xi}_{1} \mu)$ 
and $({\hat \xi}_{2} \mu)$ both belong to the family ${\cal M}$
and have degree $1$.

 It's not difficult to check the relation

$$\left[ {\hat \xi}_{2}, {\hat \xi}_{1} \right] \,\,\, = \,\,\,
{\hat \xi}_{2} \, {\hat \xi}_{1} \, - \, {\hat \xi}_{1} \,
{\hat \xi}_{2} \,\,\, = \,\,\,  {1 \over \mu} 
\left( {\hat \xi}_{2} \mu \right) {\hat \xi}_{1}
\, - \, {1 \over \mu} 
\left( {\hat \xi}_{1} \mu \right) {\hat \xi}_{2} \, + \, 
\left( S_{TT} - S_{XX} \right) {\hat \xi}_{2} \,\,\, = $$
\begin{equation}
\label{commut}
= \,\,\,   {1 \over \mu}
\left( {\hat \xi}_{2} \mu \right) {\hat \xi}_{1}
\, - \, {1 \over \mu}
\left( {\hat \xi}_{1} \mu \right) {\hat \xi}_{2} \, + \,
\left( \sum_{k\geq1} \nu_{(k)} \right) \, {\hat \xi}_{2}
\end{equation}

 We have then
\begin{equation}
\label{xi2xi1mu}
{\hat \xi}_{2} \, {\hat \xi}_{1} \, \mu \,\,\, = \,\,\,
{\hat \xi}_{1} \, {\hat \xi}_{2} \, \mu   \, + \, \sum_{k\geq1}
\left( {\hat \xi}_{2} \mu \right) \, \nu_{(k)}
\end{equation}
which gives the gradated representation of the function
${\hat \xi}_{2} \, {\hat \xi}_{1} \, \mu$.

 Let us consider now the function 
${\hat \xi}_{2} \, {\hat \xi}_{2} \, \mu \, = \, 
{\hat \xi}_{2}^{2} \, \mu$. By direct calculation the following
relation can be proved

$${\hat \xi}_{2}^{2} \mu \,\, = \,\, {\hat \xi}_{1}^{2} \mu
\, - \, 2 \mu \, {\hat \xi}_{1} \left( S_{TT} - S_{XX} \right)
\, - \, {2 \over \mu} \left( {\hat \xi}_{1} \mu \right)^{2}
\, + \, {2 \over \mu} \left( {\hat \xi}_{2} \mu \right)^{2}
\, + $$
$$+ \, 3  \left( {\hat \xi}_{1} \mu \right)
\left( S_{TT} - S_{XX} \right) \, - \, 
2 \mu \left( S_{TT} - S_{XX} \right)^{2} $$

 Writing again

$${\hat \xi}_{2}^{2} \mu \,\, = \,\, {\hat \xi}_{1}^{2} \mu
\, - \, {2 \over \mu} \left( {\hat \xi}_{1} \mu \right)^{2}
\, + \, {2 \over \mu} \left( {\hat \xi}_{2} \mu \right)^{2}
\, - \, 2 \mu \, \sum_{k\geq1} {\hat \xi}_{1} \nu_{(k)} \, + $$
\begin{equation}
\label{xi2sqmu}
+ \, 3  \left( {\hat \xi}_{1} \mu \right) 
\sum_{k\geq1} \nu_{(k)} \, - \, 
2 \mu \, \left( \sum_{k\geq1} \nu_{(k)} \right)^{2}
\end{equation}
we get the gradated representation of ${\hat \xi}_{2}^{2} \mu$.

 All the other (monomial) expressions ${\hat \xi}_{2} F$
with $deg \, F = k$ can be represented in one of the following 
two forms

1) ${\hat \xi}_{2} F \, = \, {\hat \xi}_{2} {\hat \xi}_{1}
F^{\prime} \,\,\,\,\,\,\,\, , \,\,\,\,\, deg \, F^{\prime} = k-1$;

2) ${\hat \xi}_{2} F \, = \, {\hat \xi}_{2} F_{1} F_{2}
\,\,\,\,\,\,\,\, , \,\,\,\,\, deg \, F_{1} < k, \,\,\, 
deg \, F_{2} < k$ .

 We have then 

\begin{equation}
\label{relone}
{\hat \xi}_{2} {\hat \xi}_{1} F^{\prime} \,\, = \,\,
{\hat \xi}_{1} {\hat \xi}_{2} F^{\prime} \, + \,
{1 \over \mu} \left( {\hat \xi}_{2} \mu \right) 
{\hat \xi}_{1} F^{\prime} \, - \, 
{1 \over \mu} \left( {\hat \xi}_{1} \mu \right)
{\hat \xi}_{2} F^{\prime} \, + \, \left( \sum_{k\geq1}
\nu_{(k)} \right) {\hat \xi}_{2} F^{\prime}
\end{equation}

\begin{equation}
\label{reltwo}
{\hat \xi}_{2} F_{1} F_{2} \,\, = \,\,
F_{1} {\hat \xi}_{2} F_{2} \, + \, F_{2} {\hat \xi}_{2} F_{1}
\end{equation}
so we can use the induction with respect to $ deg \, F$.

{\hfill Lemma 2.1 is proved.}

\vspace{0.5cm}

 According to Lemma 2.1 it's natural to consider now the functions
$f(\mu)$ ($f$ any smooth), $A = {\hat \xi}_{1} \mu$,
$B = {\hat \xi}_{2} \mu$, $A_{(k)} = {\hat \xi}_{1}^{k} A$,
$B_{(k)} = {\hat \xi}_{1}^{k} B$, as the "generators" for the
gradated expansions of the solutions (\ref{newexp}). We have then
according to our rule: $deg \, f(\mu) = 0$, 
$deg \, A = \deg \, B = 1$, 
$deg \, A_{(k)} = \deg \, B_{(k)} = k+1$.

 We will need also another technical Lemma:

\vspace{0.5cm}

{\bf Lemma 2.2.}

{\it For any function $F$ of degree $k$ the terms 
$({\hat \xi}_{2} F)^{[k+1]}$, $\dots$, $({\hat \xi}_{2} F)^{[k+s]}$,
$(s \geq 1)$ of the gradated expansion of ${\hat \xi}_{2} F$ are 
defined by the terms $\nu_{(1)}$, $\dots$, $\nu_{(s)}$ of the
system (\ref{defLor}).}

\vspace{0.5cm}

Proof.

Easy to see that this statement is true for
${\hat \xi}_{2} {\hat \xi}_{1} \mu$ 
($F = {\hat \xi}_{1} \mu$) and ${\hat \xi}_{2}^{2} \mu$
($F = {\hat \xi}_{2} \mu$). Using then the same induction
with respect to $deg \, F$ and the relations 
(\ref{relone})-(\ref{reltwo}) we get the statement of the
Lemma.

{\hfill Lemma 2.2 is proved.}

\vspace{0.5cm}

 Using Lemma 2.2 we can prove now the following important Lemma:

\vspace{0.5cm}

{\bf Lemma 2.3.}

{\it The solvability condition for the system (\ref{xiexpr}) 
in the $k$-th order defines uniquely the term $\nu_{(k)}$ of the
system (\ref{defLor}) provided that the terms $\nu_{(1)}$,
$\dots$, $\nu_{(k-1)}$ and corrections ${\tilde \Psi}_{(1)}$,
$\dots$, ${\tilde \Psi}_{(k-1)}$ are known.}

\vspace{0.5cm}

Proof.

Indeed, we have for $k \geq 2$

$${1 \over \mu} \left({\hat \xi}_{2}^{2} \phi \right)^{[k]}
\,\, = \,\, \sum_{l=0}^{k-2} \sum_{s=1}^{k-l-1} \left(
{\hat \xi}_{2} \left( {\hat \xi}_{2} {\tilde \Psi}_{(l)} 
\right)^{[l+s]} \right)^{[k]} $$
where $s \leq k-1$, $\,\,$ $k-l-s \leq k-1$.

 In the same way

$$\left( {\hat \xi}_{2} \phi \right)^{[k-1]} \,\, = \,\,
\sum_{l=0}^{k-2} \left( {\hat \xi}_{2} {\tilde \Psi}_{(l)} 
\right)^{[k-1]} $$
where $k-l-1 \leq k-1$.

 All the functions ${\tilde \Psi}_{(0)}$, $\dots$,
${\tilde \Psi}_{(k-2)}$ depend on $\nu_{(1)}$, $\dots$,
$\nu_{(k-2)}$ (and $(k, \omega)$). Using Lemma 2.2 we can claim
then that both the expressions above are defined by the terms
$\nu_{(1)}$, $\dots$, $\nu_{(k-1)}$.

 Looking at the other terms in the right-hand part of 
(\ref{xiexpr}) we can see that they all depend on
$\nu_{(1)}$, $\dots$, $\nu_{(k-1)}$, $\,\,$
${\tilde \Psi}_{(0)}$, $\dots$,
${\tilde \Psi}_{(k-1)}$ except one term
$\nu_{(k)} {\tilde \Psi}_{(0)\theta}$. We have then that the
orthogonality condition (\ref{ortcond}) in the order $k$ gives
the relation

$$\nu_{(k)} \int_{0}^{2\pi} \Phi_{\theta}^{2} \,
{d \theta \over 2\pi} \,\,\, = \,\,\, 
G_{(k)} \left(k, \omega, \nu_{(1)}, \dots, \nu_{(k-1)} \right)$$
where $G_{(k)}$ is some smooth functional of
$(k, \omega, \nu_{(1)}, \dots, \nu_{(k-1)})$.

{\hfill Lemma 2.3 is proved.}

\vspace{0.5cm}
 
 Let us use now the natural choice of the functions
$\Phi(\theta,\mu)$ determined by the requirement that
$\Phi(\theta,\mu)$ has a local minimum at the point $\theta=0$.
Easy to see that $\Phi(\theta,\mu)$ is a symmetric function
of $\theta$: $\Phi(\theta,\mu) = \Phi(-\theta,\mu)$ in this case.
The orthogonality condition (\ref{ortcond}) for $k=1$ gives
the formula (\ref{nu1}) for the function $\nu_{(1)}$ and we
obtain the Whitham system as the main term of the system
(\ref{defLor}). The function ${\tilde f}_{(1)}$ given by

$${\tilde f}_{(1)} \,\, = \,\, - \nu_{(1)} \Phi_{\theta} \, - \,
2 S_{T} \Phi_{\theta T} \, + \, 2 S_{X} \Phi_{\theta X} $$
is anti-symmetric in $\theta$:
${\tilde f}_{(1)}(-\theta,X,T) = - {\tilde f}_{(1)}(\theta,X,T)$.
Let us formulate now the following Lemma about the solutions 
${\tilde \Psi}_{(k)}$ of the system (\ref{newksyst})
proved in \cite{deflag}:

\vspace{0.5cm}

{\bf Lemma 2.4.}

{\it 
1) For a smooth periodic anti-symmetric function 
${\tilde f}_{(k)}(\theta)$ satisfying the orthogonality 
conditions (\ref{ortcond}) the solution 
${\tilde \Psi}_{(k)}(\theta)$ satisfying the normalization
conditions (\ref{normcond}) is a smooth periodic
anti-symmetric function: 
${\tilde \Psi}_{(k)}(-\theta) = - {\tilde \Psi}_{(k)}(\theta)$.

2) For a smooth periodic symmetric function 
${\tilde f}_{(k)}(\theta)$ the solution
${\tilde \Psi}_{(k)}(\theta)$ satisfying the normalization
conditions (\ref{normcond}) is a smooth periodic
symmetric function:
${\tilde \Psi}_{(k)}(-\theta) = {\tilde \Psi}_{(k)}(\theta)$.}
\footnote{For another normalization of the functions
$\Psi_{(k)}$ the analogous property was pointed out in 
\cite{AblBenny}.}

\vspace{0.5cm}

 We can claim then that the function 
${\tilde \Psi}_{(1)}(\theta,X,T)$ is a periodic anti-symmetric
function of $\theta$: 
${\tilde \Psi}_{(1)}(-\theta,X,T) = 
- {\tilde \Psi}_{(1)}(\theta,X,T)$.

 Easy to see now that the discrepancy function 
${\tilde f}_{(2)}$ can be represented in the form

$${\tilde f}_{(2)} \,\, = \,\, - \nu_{(2)} 
\Phi_{\theta} (\theta, \mu) \, + \,
{\tilde f}^{\prime}_{(2)}(\theta,X,T) $$
where ${\tilde f}^{\prime}_{(2)}$ is symmetric in $\theta$:
${\tilde f}^{\prime}_{(2)}(-\theta,X,T) =
{\tilde f}^{\prime}_{(2)}(\theta,X,T)$.

 Using the orthogonality conditions (\ref{ortcond}) we obtain
then: $\nu_{(2)} \equiv 0$ for the second term of the system
(\ref{defLor}). We have then 
${\tilde f}_{(2)} = {\tilde f}^{\prime}_{(2)}$ and 
${\tilde \Psi}_{(2)}$ is then a symmetric function of $\theta$:
${\tilde \Psi}_{(2)}(-\theta,X,T) = 
{\tilde \Psi}_{(2)}(\theta,X,T)$. Using the induction we obtain
the following Lemma:
 
\vspace{0.5cm}

{\bf Lemma 2.5.}

{\it Under the choice of the functions $\Phi(\theta,\mu)$
given above the following statements are true:

1) All the even terms $\nu_{(2l)}(k, \omega, \dots)$
in the deformation of Whitham system (\ref{defLor}) are   
identically zero: $\nu_{(2l)} \equiv 0$;

2) All the odd corrections ${\tilde \Psi}_{(2l+1)}(\theta,X,T)$,
$l \geq 0$ in (\ref{newexp}) are anti-symmetric in $\theta$:
${\tilde \Psi}_{(2l+1)}(-\theta) = 
- {\tilde \Psi}_{(2l+1)}(\theta)$;

3) All the even corrections ${\tilde \Psi}_{(2l)}(\theta,X,T)$,
$l \geq 1$ in (\ref{newexp}) are symmetric in $\theta$:
${\tilde \Psi}_{(2l)}(-\theta) = 
{\tilde \Psi}_{(2l)}(\theta)$. }
\footnote{The analogous Lemma is true also for the deformation
of the Whitham system having Dubrovin - Zhang form 
(\cite{deflag}).}

\vspace{0.5cm}

 Thus we can rewrite the system (\ref{defLor}) in the form

\begin{equation}
\label{def1}
S_{TT} \, - \, S_{XX} \,\,\, = \,\,\, \sum_{l\geq0} \nu_{(2l+1)}
\end{equation}
where every $\nu_{(2l+1)}$ is a sum of functions belonging to
the family ${\cal M}$ and having degree $2l+1$.\footnote{The
dispersion corrections arising here are in fact rather different
from those considered in \cite{AblBenny} because of the 
"total renormalization" of parameters used in our approach.}

\vspace{0.5cm}

 Let us point out also that the system (\ref{def1}) inherits
the momentum and the energy conservation laws of the system
(\ref{nKG}) which are given by the restriction of the
corresponding laws to the solutions (\ref{newexp}) and then
integration with respect to $\theta$. The corresponding relations
are given then by the (infinite) series depending on $k$,
$\omega$ and their derivatives which can be written according
to the gradation rule. Thus we can write

\begin{equation}
\label{mom}
\langle \varphi_{T} \varphi_{X} \rangle_{T} \,\,\, = \,\,\,
\langle {\varphi_{T}^{2} \over 2} + {\varphi_{X}^{2} \over 2} -
V(\varphi) \rangle_{X}
\end{equation}
(momentum conservation)

\begin{equation}
\label{energ}
\langle {\varphi_{T}^{2} \over 2} + {\varphi_{X}^{2} \over 2} +
V(\varphi) \rangle_{T} \,\,\, = \,\,\, 
\langle \varphi_{T} \varphi_{X} \rangle_{X}
\end{equation}
(energy conservation).

 We assume here that the function $\varphi(\theta,X,T)$ is
given by the expression (\ref{newexp}) and the notation
$< \dots >$ is used here for the "averaging" procedure given
by the integration w.r.t. $\theta$: $\,\,\,$
$< \dots > \equiv \int_{0}^{2\pi} \dots d\theta/2\pi$ .

\section{Lagrangian formalism and the deformation \newline 
procedure.} 
\setcounter{equation}{0}

 Let us use now the Lagrangian formalism of the initial system
(\ref{nKG}) to obtain the first non-trivial correction $\nu_{(3)}$
in the system (\ref{def1}). We will use here the scheme suggested 
in \cite{deflag} for the Lagrangian systems. It is well known
that the equation (\ref{nKG}) can be represented in the Lagrangian
form

$${\delta \over \delta \varphi(x,t)} \int_{-\infty}^{\infty}
\int_{-\infty}^{\infty} \left[ - {1 \over 2} \, \varphi_{t}^{2} 
\, + \, {1 \over 2} \, \varphi_{x}^{2} \, + \, V(\varphi) \right]
dx \, dt \,\,\, = \,\,\, 0 $$

 We have to add the variable $\theta$ and introduce the action
functional

\begin{equation}
\label{lagrKG}
\Sigma [\varphi] \,\,\, = \,\,\, \int\int \int_{0}^{2\pi} 
\left[ - {1 \over 2} \, \varphi_{T}^{2} \, + \,
{1 \over 2} \, \varphi_{X}^{2} \, + \, V(\varphi) \right] \,
{d \theta \over 2\pi} \, dX \, dT
\end{equation}

 In terms of the function $\phi(\theta,X,T)$ we can write
then

$$\Sigma \,\,\, = \,\,\, \int\int \int_{0}^{2\pi}
\left[ - \, {1 \over 2} \, S_{T}^{2}
\left( \phi_{\theta} \right)^{2} \, + \,
{1 \over 2} \,
S_{X}^{2} \left( \phi_{\theta} \right)^{2} \, + \,
V \left( \phi \right) \right] \,
{d \theta \over 2\pi} \, dX \, dT \,\, + $$
$$+ \,\, \int\int \int_{0}^{2\pi} \left[ - \,
S_{T} \, \phi_{\theta} \phi_{T} \, + \,
S_{X} \, \phi_{\theta} \phi_{X} \right] \,
{d \theta \over 2\pi} \, dX \, dT \,\, + $$
\begin{equation}
\label{thetact}
+ \,\, \int\int \int_{0}^{2\pi} {1 \over 2} \left[ - \,
\left( \phi_{T} \right)^{2} \, + \,
\left( \phi_{X} \right)^{2} \right] \,
{d \theta \over 2\pi} \, dX \, dT
\end{equation}

 The system (\ref{def1}) is equivalent to the equation

$${\delta \Sigma \over \delta S(X,T)} \,\,\, = \,\,\, 0$$
where the function $\phi(\theta,X,T)$ is given by the relation
(\ref{phifunc}).

 For the determination of $\nu_{(3)}$ we need to write 
just the part $\Sigma^{\prime} = \Sigma_{(0)} + \Sigma_{(2)}$
of the action (\ref{thetact}), where

$$\Sigma_{(0)} \,\,\, = \,\,\, \int\!\!\int\!\!\int_{0}^{2\pi}
\left( - \, {1 \over 2} \mu \Phi_{\theta}^{2} \, + \,
V(\Phi) \right) {d \theta \over 2\pi} \, dX \, dT $$
and

$$\Sigma_{(2)} \,\,\, = \,\,\, {1 \over 2} 
\int\!\!\int\!\!\int_{0}^{2\pi}
\left( - \mu {\tilde \Psi}_{(1)\theta}^{2} \, + \,
V^{\prime\prime}(\Phi) {\tilde \Psi}_{(1)}^{2} \right) 
{d \theta \over 2\pi} \, dX \, dT \, +$$
$$+ \, \int\!\!\int\!\!\int_{0}^{2\pi} \left(
2 S_{T} \, \Phi_{\theta T} \, - \, 2 S_{X} \, \Phi_{\theta X}
\right) {\tilde \Psi}_{(1)} \, {d \theta \over 2\pi} \, dX \, dT 
\, +$$
$$+ \, {1 \over 2} \int\!\!\int\!\!\int_{0}^{2\pi}
\left( - \Phi_{T}^{2} \, + \, \Phi_{X}^{2} \right)
{d \theta \over 2\pi} \, dX \, dT $$
or, equivalently:  

$$\Sigma_{(2)} \,\, = \,\, \int\!\!\int {1 \over 2\mu} 
\langle \Phi_{\mu}^{2} \rangle
\left({\hat \xi}_{2} \mu \right)^{2} \, dX \, dT \, +$$
$$+ \, \int\!\!\int \left[
- {1 \over 2} \mu \langle Z^{2}(\theta,\mu) \rangle +
{1 \over 2} \langle V^{\prime\prime}(\Phi) Z^{2}(\theta,\mu) 
\rangle + 2 \langle \Phi_{\theta\mu} Z (\theta,\mu) \rangle 
- {1 \over 2\mu} \langle \Phi_{\mu}^{2} \rangle \right]
\left({\hat \xi}_{1} \mu \right)^{2} \, dX \, dT $$
where the function $Z(\theta,\mu)$ is defined from the 
representation of ${\tilde \Psi}_{(1)}$ in the form

$${\tilde \Psi}_{(1)} (\theta, X, T) \,\,\, = \,\,\,
Z(\theta,\mu) \, \left({\hat \xi}_{1} \mu \right) $$
after the resolving of the system (\ref{newksyst}) for $k=1$
with the normalization conditions (\ref{normcond}).

 (Let us note that

$${\tilde f}_{(1)} \,\, = \,\, - \nu_{(1)} \Phi_{\theta} 
\, - \, 2 {\hat \xi}_{1} \Phi_{\theta} \,\, = \,\,
\left( {W^{\prime}(\mu) \over W(\mu)} \Phi_{\theta} \, - \, 
2 \Phi_{\theta\mu} \right) \left({\hat \xi}_{1} \mu \right) $$
for $k=1$.)

 It can be shown after some calculations that

$${\delta \Sigma_{(2)} \over \delta S(X,T)} \,\,\, = \,\,\,
\left[ - \left( S_{TT} - S_{XX} + {\hat \xi}_{1} \right)
\left({\hat \xi}_{2} \mu \right)^{2} {d \over d\mu} \, + \,
{\hat {\cal O}}_{2} \left({\hat \xi}_{2} \mu \right) \right]
{1 \over \mu} \langle \Phi_{\mu}^{2} \rangle \, + $$

$$+ \, \left[ -2 \left( S_{TT} - S_{XX} + {\hat \xi}_{1} \right) 
\left({\hat \xi}_{1} \mu \right)^{2} {d \over d\mu} \, + \,
2 {\hat {\cal O}}_{1} \left({\hat \xi}_{1} \mu \right) \right]
\, \times $$
$$\times \, \left[
- {1 \over 2} \mu \langle Z^{2}(\theta,\mu) \rangle +
{1 \over 2} \langle V^{\prime\prime}(\Phi) Z^{2}(\theta,\mu)
\rangle + 2 \langle \Phi_{\theta\mu} Z (\theta,\mu) \rangle
- {1 \over 2\mu} \langle \Phi_{\mu}^{2} \rangle \right] $$
where the operators ${\hat {\cal O}}_{1}$ and 
${\hat {\cal O}}_{2}$ are given by the expressions

$${\hat {\cal O}}_{1} \,\, = \,\, 2 {\hat \xi}_{1}^{2} \, + \,
\left( 4 \left( S_{TT} - S_{XX} \right) \, - \, {1 \over \mu} 
\left({\hat \xi}_{1} \mu \right) \right) {\hat \xi}_{1} \, + \,
{1 \over \mu} \left({\hat \xi}_{2} \mu \right) {\hat \xi}_{2} 
\, - $$
$$- \, {1 \over \mu^{2}} \left({\hat \xi}_{1} \mu \right)^{2}
\, + \, {1 \over \mu^{2}} \left({\hat \xi}_{2} \mu \right)^{2}
\, + \, {2 \over \mu} \left({\hat \xi}_{1} \mu \right)
\left( S_{TT} - S_{XX} \right) $$

$${\hat {\cal O}}_{2} \,\, = \,\, 
2 {\hat \xi}_{1} {\hat \xi}_{2} \, - \, 
{1 \over \mu} \left({\hat \xi}_{2} \mu \right) {\hat \xi}_{1}
\, + \, {1 \over \mu} \left({\hat \xi}_{1} \mu \right)
{\hat \xi}_{2} \, + \, 2 \left( S_{TT} - S_{XX} \right)
{\hat \xi}_{2} $$

 To obtain the function $\nu_{(3)}$ we need in fact just the main 
term (of degree $3$) of the gradated expansion of this expression.
To get this term we have to change first of all the expression
$S_{TT} - S_{XX}$ by it's main terms $\nu_{(1)}$ in the gradated
expansion. Besides that, we have to take the main terms in the
gradated expansions of the expressions 
${\hat \xi}_{2} {\hat \xi}_{1} \mu $ and
${\hat \xi}_{2} {\hat \xi}_{2} \mu $ arising in the expression
above. Easy to see that these terms can be just extracted from
the formulae (\ref{xi2xi1mu}) and (\ref{xi2sqmu}) and are equal
to

$${\hat \xi}_{1} {\hat \xi}_{2} \mu \, + \, 
\nu_{(1)} \, {\hat \xi}_{2} \mu $$
and

$${\hat \xi}_{1}^{2} \mu \, - \, 
{2 \over \mu} \left( {\hat \xi}_{1} \mu \right)^{2} \, + \,
{2 \over \mu} \left( {\hat \xi}_{2} \mu \right)^{2} \, - \,
2 \mu \, {\hat \xi}_{1} \nu_{(1)} \, + \,
3 \left( {\hat \xi}_{1} \mu \right) \nu_{(1)} \, - \,
2 \mu \, \nu_{(1)}^{2} $$
respectively. 

 The expression $\delta \Sigma_{(0)}/\delta S(X,T)$ is equal to

$$\left( S_{TT} - S_{XX} \right) W(\mu) \, + \,
{\hat \xi}_{1} \, W(\mu) $$
so we have

$$\left( {\delta \Sigma_{(0)} \over \delta S(X,T)} \right)^{[3]}
\,\,\, = \,\,\, \nu_{(3)} \, W(\mu) $$
(and the Whitham system in the order $1$).

 Finally we have for $\nu_{(3)}$:

$$\nu_{(3)} \,\,\, = \,\,\, - {1 \over W(\mu)}
\left( {\delta \Sigma_{(2)} \over \delta S(X,T)} \right)^{[3]} $$

\vspace{0.5cm}

 The work was partially supported by the grant of President of 
Russian Federation (MD-8906.2006.2) and Russian Science Support
Foundation.

\end{document}